\documentclass[cameraready]{Interspeech}
\usepackage{multirow}
\usepackage{graphicx}
\usepackage{booktabs}
\usepackage{caption}
\usepackage{lscape}
\usepackage{tabularx}
\usepackage{xcolor}
\usepackage{amssymb}
\usepackage{amsmath,graphicx,hyperref}
\usepackage{pifont}
\usepackage{makecell}
\usepackage{array}
\usepackage{graphicx}
\usepackage{subcaption}

\title{Aligning Audio Captions with Human Preferences}

\author[affiliation={1}]{Kartik}{Hegde}
\author[affiliation={2}]{Rehana}{Mahfuz}
\author[affiliation={2}]{Yinyi}{Guo}
\author[affiliation={2}]{Erik}{Visser}

\address{
    $^1$ Qualcomm Technologies, Inc., India \\
    $^2$ Qualcomm Technologies, Inc., USA
}
\email{\{karthegd, rmahfuz, yinyig, evisser\}@qti.qualcomm.com}


\usepackage{comment}

\begin{document}

\maketitle

\begin{abstract}
    Current audio captioning relies on supervised learning with paired audio-caption data, which is costly to curate and may not reflect human preferences in real-world scenarios. To address this, we propose a preference-aligned audio captioning framework based on Reinforcement Learning from Human Feedback (RLHF). To capture nuanced preferences, we train a Contrastive Language-Audio Pretraining (CLAP) based reward model using human-labeled pairwise preference data. This reward model is integrated into an RL framework to fine-tune any baseline captioning system without ground-truth annotations. Extensive human evaluations across multiple datasets show that our method produces captions preferred over baseline models, particularly when baselines fail to provide correct and natural captions. Furthermore, our framework achieves performance comparable to supervised approaches with ground-truth data, demonstrating effective alignment with human preferences and scalability in real-world use.
\end{abstract}

\section{Introduction}
\label{sec:intro}

Audio captioning, the task of generating natural language descriptions for acoustic scenes, has become an important benchmark for evaluating audio-language models. Unlike image captioning, which benefits from explicit spatial and semantic cues, audio captioning faces inherent ambiguity and temporal complexity, making accurate and human-aligned caption generation particularly challenging.

While supervised approaches using datasets such as AudioCaps \cite{audiocaps} and Clotho \cite{clotho} have advanced the field, their reliance on automatic metrics like BLEU \cite{bleu}, METEOR \cite{meteor}, and CIDEr \cite{cider} introduces limitations, as these metrics correlate poorly with human judgments \cite{fense}. Moreover, recent contrastive learning methods such as CLAP, though scalable, fail to capture nuanced human preferences, underscoring the need for approaches that optimize for human alignment rather than static similarity metrics. 
Optimizing task-specific metrics using Reinforcement Learning (RL) was popularized in image captioning through Self-Critical Sequence Training (SCST) \cite{scst}. However, metric-driven RL optimization remains relatively under explored in audio captioning. One possible reason is that the same acoustic scene can often be described at different semantic granularities by different listeners, further complicating reliable supervision and evaluation. Recent work \cite{optimizingCLAP} has shown that using a CLAP-based reward model in combination with a large language model can improve alignment with human expectations, outperforming traditional CIDEr-based RL approaches on mean-opinion-score (MOS) human evaluations. Despite these improvements, the dependence on large external language models introduces substantial computational overhead, making such methods impractical for deployment in low‑memory, resource‑constrained audio captioning systems. Moreover, evaluating audio captions using absolute ratings often leads to poor inter‑annotator agreement \cite{fense}, further motivating preference-based approaches. 

To address the challenges associated with audio captioning such as the reliance on paired audio-caption ground truth data and the limitations of existing evaluation metrics we propose a framework that enhances caption quality based on human preferences. Our method does not require additional paired data, making it more scalable and cost-effective. Since collecting preference data is significantly easier and less resource-intensive than obtaining ground-truth captions, we design a simple yet effective reward model that leverages pairwise human judgments. To ensure training stability and mitigate reward hacking during reinforcement learning, we incorporate reward shaping techniques \cite{rewardshaping}. We conduct a comparative study between captions generated by the baseline model and those produced by our proposed method, evaluated both through human assessments and established automatic metrics. Additionally, we analyze the performance of our approach relative to supervised training methods.


Our key contributions are:

\begin{itemize}
\item We propose a novel audio captioning framework that optimizes a custom reward function using reinforcement learning, without requiring paired audio-caption data.
\item We develop a CLAP embedding-based reward model trained on pairwise human preference data to effectively capture subjective judgments.
\item We conduct extensive human evaluations on publicly available and proprietary datasets, showing that our RLHF-based fine-tuning outperforms baseline models, especially when the baseline model generated captions are incorrect or unnatural. Additionally, our method achieves performance comparable to supervised training, while requiring no labelled audio-caption data.
\end{itemize}


\begin{figure}[!t]
    \centering
    \includegraphics[width=0.95\columnwidth]{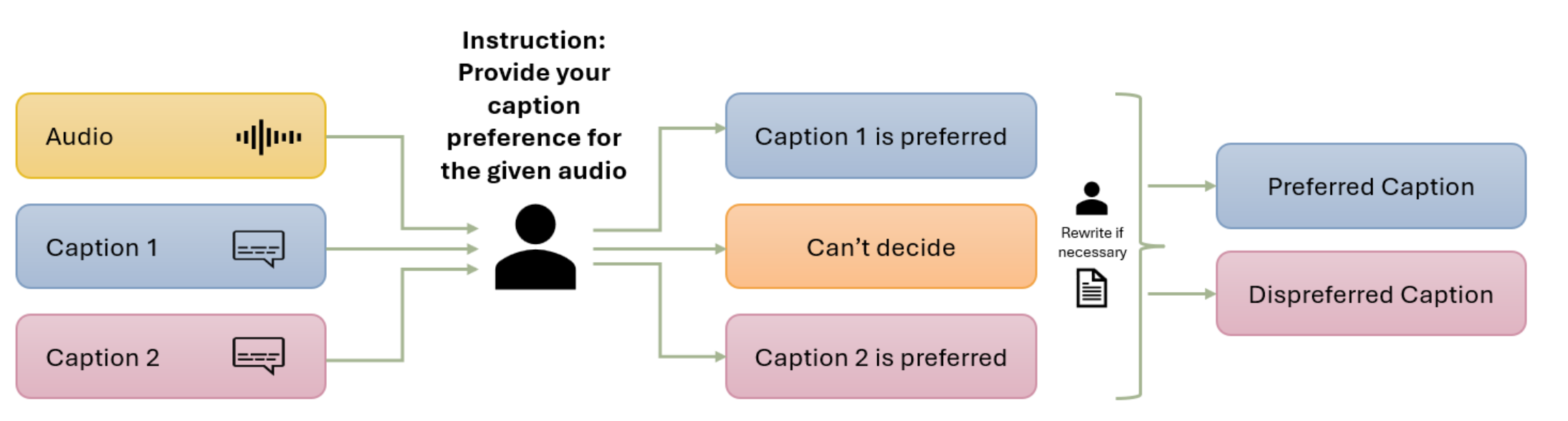}
    \caption{Pairwise human preference annotation and subjective evaluation setup}
    \label{fig:pref_figure}
\end{figure}

\begin{figure*}[!t]
    \centering
    \includegraphics[width=0.8\textwidth]{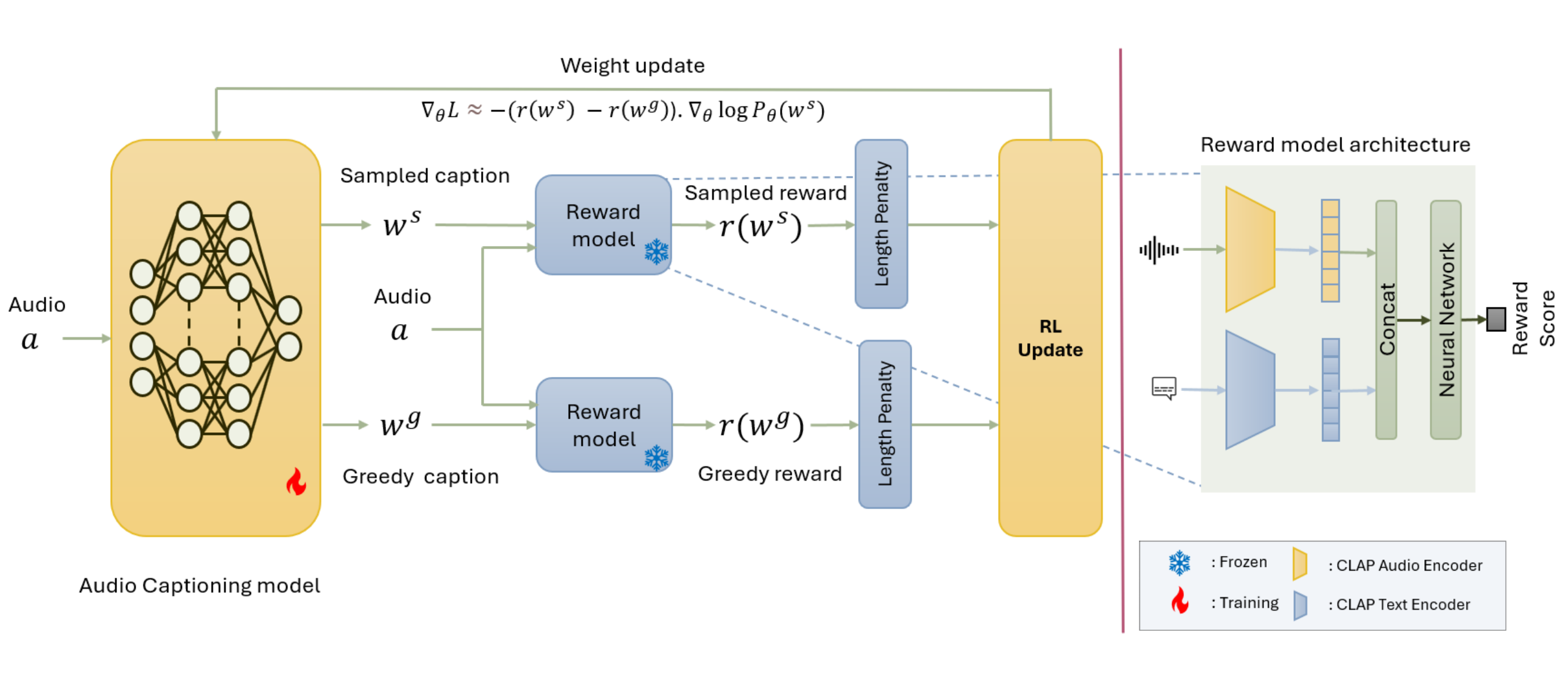}
    \caption{Proposed method for RL using custom reward model (left) and CLAP based architecture of the custom reward model inference(right).}
    \label{fig:AudioCap_combined_fig}
\end{figure*}


\section{Related Work}
\label{sec:related}

Early works in audio captioning, such as AudioCaps \cite{audiocaps} and Clotho \cite{clotho}, introduced large-scale datasets pairing audio clips with human-written captions, enabling supervised fine-tuning (SFT) of captioning models. Evaluation has traditionally relied on image captioning metrics like BLEU \cite{bleu}, METEOR \cite{meteor}, and CIDEr \cite{cider}, but these metrics poorly capture semantic fidelity and subjective quality \cite{fense,detecting_false_alarm}. Alternatives such as FENSE \cite{fense, mace} combine semantic similarity with error modeling for improved human alignment.

CLAP-based models \cite{msclap,laionclap} learn joint embeddings for audio and text, and CLAPScore has been used as a proxy for caption relevance and as a hallucination metric \cite{parameter}. However, recent studies indicate that CLAPScore correlates only weakly with human judgments \cite{humanclap}, and that CLAP embeddings suffer from a modality gap that limits fine-grained semantic alignment \cite{modalitygap}. These findings suggest that optimizing solely for CLAP similarity may produce captions that are technically relevant but lack naturalness and contextual depth. To address this, we adapt a neural network over CLAP embeddings to better capture human preferences.

RLHF has emerged as a powerful paradigm for aligning model outputs with human expectations, particularly in large language models. Self-Critical Sequence Training (SCST) \cite{scst} applies policy gradient methods to optimize non-differentiable metrics like CIDEr. Alternative approaches such as Direct Preference Optimization (DPO) \cite{dpo} have shown promise in text domains but, due to reliance on static pairwise comparisons and limited control over output structure \cite{ppovsdpo}, they are less suited for concise audio captioning systems. Our work builds on SCST and extends RLHF to audio captioning with a custom reward model and reward shaping strategies to stabilize training \cite{optimizingCLAP, rewardshaping}.

\begin{table*}[ht]
\centering
\caption{Metric-wise comparison of baseline and proposed RLHF systems}
\label{tab:proprietary_pref_comparison}
\scriptsize

\begin{tabularx}{\textwidth}{
l l l
>{\centering\arraybackslash}X
>{\centering\arraybackslash}X
>{\centering\arraybackslash}X
>{\centering\arraybackslash}X
>{\centering\arraybackslash}X
>{\centering\arraybackslash}X
>{\centering\arraybackslash}X
>{\centering\arraybackslash}X
}
\toprule

\textbf{Dataset} & \textbf{Test Data} & \textbf{System} &
\multicolumn{2}{c}{\textbf{N-gram}} &
\multicolumn{3}{c}{\textbf{Semantic}} &
\textbf{Custom} &
\multicolumn{2}{c}{\shortstack{\textbf{Proposed RLHF}\\\textbf{Win Rate (\%)}}} \\

\cmidrule(lr){4-5} \cmidrule(lr){6-8} \cmidrule(lr){9-9} \cmidrule(lr){10-11}

& & &
\textbf{BLEU$_4$↑} & \textbf{CIDEr↑} &
\textbf{S-BERT↑} & \textbf{FENSE↑} & \textbf{CLAP\textsubscript{AT}↑} &
\textbf{Reward↑} &
\textbf{Reward↑} & \textbf{Human↑} \\

\midrule

AudioCaps & Non-challenging & Baseline
& 0.0272 & \textbf{0.3211} & \textbf{0.5947} & \textbf{0.5947} & \textbf{0.3517}
& \textbf{0.6524} & 48.20 & \textbf{52.03} \\

AudioCaps & Non-challenging & +RLHF
& \textbf{0.0314} & 0.3178 & 0.5836 & 0.5282 & 0.3231
& 0.6480 & \textbf{51.80} & 47.97 \\
\midrule

AudioCaps & Challenging & Baseline
& \textbf{0.0541} & \textbf{0.3357} & 0.4565 & 0.1441 & 0.2546
& 0.5482 & 40.86 & 46.07 \\

AudioCaps & Challenging & +RLHF
& 0.0498 & 0.3465 & \textbf{0.5007} & \textbf{0.3571} & \textbf{0.2572}
& \textbf{0.6025} & \textbf{59.14} & \textbf{53.93} \\
\midrule

Proprietary & Non-challenging & Baseline
& \textbf{0.0919} & \textbf{0.7970} & \textbf{0.5392} & \textbf{0.4772} & \textbf{0.3015}
& 0.1958 & 35.49 & 46.59 \\

Proprietary & Non-challenging & +RLHF
& 0.0771 & 0.6324 & 0.5371 & 0.4688 & 0.2820
& \textbf{0.2981} & \textbf{64.51} & \textbf{53.41} \\
\midrule

Proprietary & Challenging & Baseline
& \textbf{0.0780} & \textbf{0.4570} & 0.5194 & 0.4637 & \textbf{0.2425}
& 0.1558 & 30.12 & 40.89 \\

Proprietary & Challenging & +RLHF
& 0.0555 & 0.4463 & \textbf{0.5607} & \textbf{0.4933} & 0.2282
& \textbf{0.3128} & \textbf{69.88} & \textbf{59.11} \\
\bottomrule
\end{tabularx}
\end{table*}

\section{Methodology}
\label{sec:methodology}

In this section we describe reward modelling for custom preference data and fine-tuning the Audio Captioning model with reinforcement learning. An overview of our proposed approach is shown in Figure~\ref{fig:AudioCap_combined_fig}.

\subsection{Reward Modelling}
\label{ssec:rm}

The reward model aims to capture subtleties in human preference data by assigning a scalar value estimating the probability of a human preferring a given audio--caption pair. Unlike reward models for text-only tasks, which typically use a pretrained language model to generate multiple completions for a prompt, our setting requires a multimodal reward model. We therefore use the CLAP audio and text encoders \cite{laionclap} to obtain 512-dimensional embeddings for each modality. These are concatenated into a 1024-dimensional joint representation and passed through a two-layer neural network with hidden sizes 512 and 128. The final output layer is a single sigmoid-activated neuron producing a value in $[0,1]$.

The preference dataset consists of tuples $(x, y_1, y_2, p)$, where $x$ is the audio input, $y_1$ and $y_2$ are candidate captions, and $p$ denotes the human-annotated preference. Following the Bradley--Terry model \cite{bt}, the probability that caption $y_1$ is preferred over $y_2$ for a given audio input $x$ is:
\begin{align}
P(y_1 \succ y_2 \mid x)
&= \frac{e^{r(x, y_1)}}{e^{r(x, y_1)} + e^{r(x, y_2)}} \notag \\
&= \sigma(r(x, y_1) - r(x, y_2)),
\label{eq:preference_model}
\end{align}
where $r(x,y)$ denotes the assigned reward and $\sigma$ is the sigmoid function.

The reward model uses a siamese architecture to compute rewards $r_{\theta}(x,y_w)$ and $r_{\theta}(x,y_l)$ for the preferred caption $y_w$ and the dispreferred caption $y_l$. Their difference is scaled by a factor $\beta$. To reduce overfitting and promote stable reward estimates, we include an L2 regularization term with coefficient $\lambda$, giving the total loss:
\begin{equation}
\mathcal{L}(\theta) = \mathcal{L}_{\text{BT}}(\theta) + \mathcal{L}_{\text{reg}}(\theta),
\label{eq:total_loss}
\end{equation}
\begin{equation}
\mathcal{L}_{\text{BT}}(\theta) = -\log \left[ \sigma \left( \beta ( r_{\theta}(x, y_w) - r_{\theta}(x, y_l) ) \right) \right],
\label{eq:bt_loss}
\end{equation}
\begin{equation}
\mathcal{L}_{\text{reg}}(\theta) = \lambda \left( r_{\theta}(x, y_w)^2 + r_{\theta}(x, y_l)^2 \right).
\label{eq:reg_loss}
\end{equation}

During inference, as shown in Figure~\ref{fig:AudioCap_combined_fig}, the reward model processes an audio--caption pair and outputs a score in $[0,1]$, serving as a proxy for the probability of human preference.

\subsection{Reinforcement Learning}
\label{ssec:rlhf}

Our objective is to enhance the baseline audio captioning model to generate captions that are more closely aligned with human preferences. We formulate this as a reward optimization problem, where the reward predicted by our reward model is maximized to obtain improved captions. Since the scalar reward is non-differentiable and direct gradient computation and model updates are not feasible, we employ the Policy Gradient method using REINFORCE \cite{REINFORCE}. To further reduce the variance of the gradient estimate, a baseline term is introduced, which does not affect the expected gradient. We employ Self-Critical Sequence Training \cite{scst} approach, which uses the output of the test-time inference algorithm, specifically, greedy decoding, as the baseline.

To compute the loss, we consider reward scores for two captions generated by the baseline model using different decoding methods. Using our custom reward calculator not only provides more control than using the traditional metrics such as CIDEr, it also eliminates the need for ground truth captions to compute the reward score, hence affording scalability. The captions $w^s$ (decoded using multinomial sampling) and $w^g$ (decoded using greedy decoding) along with the audio are passed to the reward model to obtain rewards, denoted as $r(w^s)$ and $r(w^g)$. These reward values are then used to compute the loss and estimate the gradient, given by:
\begin{equation}
\nabla_{\theta} L(\theta) \approx -(r(w^s) - r(w^g)) \nabla_{\theta} \log p_{\theta}(w^s)
\label{eq:scst_loss}
\end{equation}
A known challenge in reinforcement learning (RL) for text generation is reward hacking, where models learn to exploit the reward function in ways such as generating excessively long captions or sometimes repeating certain words that superficially maximize the reward without improving the semantic quality or relevance.

\subsubsection{Length Penalty}
\label{ssec:lenpen}

During our initial experiments, we observed that length based reward hacking was dominant. To address reward hacking, we apply reward shaping by introducing a length penalty aligning with recent work by \cite{rewardshaping}, who demonstrate how reward shaping can mitigate reward hacking. We define the modified reward function as:
\begin{equation}
r_{\text{new}} = r_{\text{old}} 
- \alpha \cdot \left( 1 - \frac{L_e}{L_c} \right) 
\cdot \max(0, L_c - L_e)
\label{eq:len_penalty}
\end{equation}

Here, $L_e$ is the expected caption length, $L_c$ is the actual generated caption length, and $\alpha$ is a tunable hyperparameter controlling the strength of the penalty. The penalty gets applied only when the generated caption length exceeds the expected length of caption. This formulation when used in equation \ref{eq:scst_loss}, penalizes overly long captions proportionally to their deviation from the expected length, helping to align the generation with user preferences.


\section{Experimental Setup}
\label{sec:expt}

We conducted our experiments on publicly available and proprietary datasets to evaluate the performance of proposed system.

\begin{table*}[!ht]
\centering
\scriptsize
\caption{Qualitative comparison of Baseline, RLHF, SFT, and human labelled Ground Truth captions on AudioCaps dataset}
\label{table:quality}
\begin{tabular}{|>{\raggedright\arraybackslash}p{3.8cm}|
                >{\raggedright\arraybackslash}p{3.9cm}|
                >{\raggedright\arraybackslash}p{2.8cm}|
                >{\raggedright\arraybackslash}p{4.6cm}|}
\hline
\multicolumn{1}{|c|}{\textbf{\rule{0pt}{15pt}Baseline}} &
\multicolumn{1}{c|}{\textbf{RLHF}} &
\multicolumn{1}{c|}{\textbf{SFT}} &
\multicolumn{1}{c|}{\textbf{Ground Truth}} \\
\hline

\multirow{3}{*}{\parbox[t]{4.0cm}{A musical genre is playing}} &
\multirow{3}{*}{\parbox[t]{4.0cm}{A musical instrument is playing}} &
\multirow{3}{*}{\parbox[t]{3.0cm}{Music plays}} &
- A piano playing as plastic bonks \\
& & & - A piano playing as a clock ticks \\
\hline

\multirow{3}{*}{\parbox[t]{4.0cm}{A man is talking while a person is snoring}} &
\multirow{3}{*}{\parbox[t]{4.0cm}{A man is talking while there is a continuous buzzing of an insect}} &
\multirow{3}{*}{\parbox[t]{3.0cm}{A man speaks and a woman laughs}} &
- Swarm of bees buzzing and two men speaking \\
& & & - Men speaking with insects buzzing \\
\hline

\end{tabular}
\end{table*}

\subsection{Dataset}
\label{ssec:ds}

To train the reward model, we construct a preference dataset containing audio samples, caption pairs, and human‑annotated choices based on caption correctness and naturalness.
We use AudioCapsEval and ClothoEval from FENSE~\cite{fense}, derived from AudioCaps~\cite{audiocaps} and Clotho~\cite{clotho}. Each contains $\sim$1,750 annotations rated by four annotators. We retain only unanimous cases, yielding 1,473 pairs from AudioCapsEval and 1,555 from ClothoEval, split 80/20 for training and validation.
To support deployment on proprietary systems, we curate 4,424 additional preference samples and 880 challenging cases. For each audio clip, captions are generated via greedy and top‑$k$ decoding, and annotators select the preferred caption or provide a correction. Samples where both captions fail are labeled challenging. As public datasets lack such labels, we approximate challenging AudioCapsEval samples by selecting the 250 lowest‑scoring clips using FENSE~\cite{fense}.
To increase diversity and better capture misalignment, we further augment with synthetic preference pairs by pairing each audio clip with a randomly chosen, unrelated caption. These negative examples help the reward model learn to penalize semantically mismatched outputs.

\begin{table}[ht!]
\centering
\caption{Comparison of Reward Models}
\label{table:reward_models}
\scriptsize
\begin{tabularx}{\columnwidth}{X c c}
\toprule
\makecell{\textbf{Reward Model Data}} &
\makecell{\textbf{Training}\\\textbf{Data Size}} &
\makecell{\textbf{Human Win Rate}\\\textbf{for RLHF}} \\
\midrule
AudioCapsEval & 1178 & 43.55\% \\
AudioCapsEval + ClothoEval & 2422 & 47.97\% \\
\bottomrule
\end{tabularx}
\end{table}


\subsection{Training and Implementation Details}
\label{ssec:training}
We adopt the model from \cite{improveacap} as our baseline audio captioning system. It employs a CNN10 PANN encoder \cite{cnn10} and a Transformer decoder, trained on 80k proprietary audio–caption pairs. With approximately 15M parameters, the model is lightweight and suitable for deployment on low‑resource or embedded devices. As limited comparably sized model built upon the DCASE baseline \cite{dcase_baseline} is publicly available, all our experiments use \cite{improveacap} as the baseline architecture.

For constructing the reward model, we extract audio and text embeddings using the pretrained LAION CLAP (htsat‑unfused) model \cite{laionclap}. The reward model is trained for 70 epochs with a batch size of 64 and a learning rate of $10^{-4}$, using output clamping to [0.01, 0.99], an L2 penalty of 0.1, and a reward‑difference scaling factor $\beta = 5$.
RLHF is performed for 100 epochs with Adam (lr $10^{-6}$, weight decay $10^{-6}$) and a batch size of 128. A two‑epoch linear warm‑up is followed by a step‑decay (×0.1 every 10 epochs). Length penalties of 0.4 (AudioCaps) and 1.0 (proprietary) are used with an expected caption length of $L_e = 13$.

\subsection{Human Evaluation}
\label{ssec:humaneval}

We collect human preference judgments following the protocol of \cite{fense}. Annotators listened to each audio sample and selected the preferred caption from two candidates, with ordering randomized to avoid bias. For comparing RLHF with the baseline, four annotators evaluated 250 caption pairs, while a smaller study with two annotators assessed 100 RLHF–SFT pairs due to annotation effort. For proprietary evaluations, seven annotators each rated 30 non‑overlapping samples from the non‑challenging and challenging splits to maximize coverage, and three annotators rated 30 overlapping samples per split to measure agreement. Inter‑annotator consistency was computed using Fleiss’ Kappa \cite{fleiss}, yielding 0.445/0.435 on AudioCaps and 0.406/0.417 on the proprietary dataset (non‑challenging/challenging), indicating acceptable level of agreement. A win is recorded when the RLHF caption is preferred, and the win rate is the percentage of RLHF wins among non‑tie cases.

\begin{table}[t]
\centering
\scriptsize

\caption{Win rate (\%) for RLHF, evaluated on AudioCaps Non-Challenging (NC) 
and Challenging (C) splits with per metric weighted deviation from 
Human preference win rates}
\label{tab:audiocaps-winrate}

\begin{tabularx}{\columnwidth}{
    >{\raggedright\arraybackslash}X 
    *{2}{>{\centering\arraybackslash}X} 
    >{\centering\arraybackslash}X}
\toprule
Metric & NC (n=707) & C (n=250) & Weighted Deviation ↓ \\
\midrule
Human  & 47.97 & 53.93 & - \\
S-BERT  & 50.00 & 64.85 & 4.35 \\
FENSE  & 43.24 & 85.64 & 11.78 \\
CLAP\textsubscript{AT}   & 42.71 & 47.57 & 5.55 \\
Reward (Ours) & 51.80 & 59.14 & \textbf{4.19} \\
\bottomrule
\end{tabularx}
\end{table}

\subsection{Pairwise comparison with different metrics}
\label{ssec:paircomp}
We evaluated how well different metrics align with human judgments using pairwise comparisons on the AudioCaps test set. The dataset was divided into Non-Challenging (NC) and Challenging (C) subsets, and we computed the win rates between the baseline and our RLHF-trained model. Table~\ref{tab:audiocaps-winrate} reports these win rates together with a weighted deviation score that quantifies how closely each metric follows human preferences in the two splits.

Our custom reward model achieves the lowest weighted deviation (\textit{4.19}), indicating the closest alignment with human evaluations among all automatic metrics. This supports our argument that a custom preference-based reward metric is effective in aligning audio captions with human preferences.

\section{Results and Discussion}
\label{sec:res}

We evaluate our RLHF framework on both public and proprietary datasets, including challenging cases where baseline models fail to produce accurate or natural captions. We observe that higher scores on automated metrics do not necessarily translate to improved human satisfaction. Table \ref{tab:proprietary_pref_comparison} shows that although the baseline performs well on metrics such as BLEU and CIDEr, it does not align with human preferences, while semantic metrics such as S-BERT and FENSE exhibit inconsistent trends, highlighting the limitations of rule-based and semantic evaluation metrics.
In contrast, our RLHF model improves caption quality, as reflected by higher human preference win rates and higher scores from our custom reward model, except on the AudioCaps evaluation set, likely due to limited preference data in that domain. Notably, RLHF yields its largest gains on challenging samples, indicating its effectiveness in correcting failure cases rather than providing marginal improvements on well-captioned examples.
Table~\ref{table:reward_models} further shows that increasing the amount of preference data improves the reward model, which in turn enhances RL fine-tuning and results in captions that better align with human preferences. With sufficient preference data, a stronger reward model can be learned to further improve state-of-the-art audio captioning systems. The preference-based reward model correlates well with human judgments, while RLHF maintains competitive performance on automatic metrics. Despite being built on $CLAP$, our reward model shows more consistent alignment with human evaluations than the conventional $CLAP_{AT}$ score.
Human win-rate analysis further indicates that low-quality baseline captions receive lower reward scores, guiding the model toward outputs that better reflect human judgment. Finally, Table~\ref{table:quality} shows examples where RLHF-generated captions align more closely with ground truth in both naturalness and correctness.

\section{Conclusions}

\label{ssec:rdiscussions}
We presented a reinforcement learning framework for audio captioning that aligns model outputs with human preferences using a custom reward model trained on preference data. Our RLHF-based approach improves human preference alignment without relying on ground-truth captions, while maintaining competitive performance on traditional metrics. The method is scalable, simple to implement, and effective in challenging scenarios, with further gains possible through enhanced reward modeling with expanded preference data and advanced RL techniques.


\bibliographystyle{IEEEtran}
\bibliography{mybib}

\end{document}